\documentclass[prl, twocolumn, superscriptaddress]{revtex4-1}

\usepackage{graphicx}
\usepackage{natbib}
\usepackage{wasysym}

\begin{document}

\title{Field-induced negative differential spin lifetime in silicon}
\author{Jing Li}
\affiliation{Department of Physics and Center for Nanophysics and Advanced Materials, University of Maryland, College Park, MD 20742, USA}
\author{Lan Qing}
\affiliation{Department of Physics and Astronomy, University of Rochester, Rochester, NY 14627}
\author{Hanan Dery}
\affiliation{Department of Physics and Astronomy, University of Rochester, Rochester, NY 14627}
\affiliation{Department of Electrical and Computer Engineering, University of Rochester, Rochester, NY 14627}
\author{Ian Appelbaum}
\altaffiliation{appelbaum@physics.umd.edu}
\affiliation{Department of Physics and Center for Nanophysics and Advanced Materials, University of Maryland, College Park, MD 20742, USA}

\begin{abstract}
We show that the electric field-induced thermal asymmetry between the electron and lattice systems in pure silicon substantially impacts the identity of the dominant spin relaxation mechanism. Comparison of empirical results from long-distance spin transport devices with detailed Monte-Carlo simulations confirms a strong spin depolarization beyond what is expected from the standard Elliott-Yafet theory already at low temperatures. The enhanced spin-flip mechanism is attributed to phonon emission processes during which electrons are scattered between conduction band valleys that reside on different crystal axes.  This leads to anomalous behavior, where (beyond a critical field) reduction of the transit time between spin-injector and spin-detector is accompanied by a counterintuitive reduction in spin polarization and an apparent \emph{negative} spin lifetime.
\end{abstract}

\maketitle

In compound semiconductors, the eventual reduction in drift velocity of conduction electrons with increasing applied electric field is known as negative differential mobility or the Gunn effect \cite{Gunn_IBM64,Kroemer_IEEE64}. In this field regime (typically several kV/cm), hot electrons scatter into low-lying secondary energy minima in the conduction band where the effective mass is larger, reducing their kinetic energy. The multivalley band structure of silicon also allows for the existence of this phenomenon but only at low temperatures; for all $T>$30~K, the drift velocity increases and eventually saturates with increasing applied field \footnote{In $T\lesssim$30~K and at electric fields around 100~V/cm, electrons predominantly scatter into valleys that are parallel to the field direction (where the effective mass is heavier). In larger fields or at higher temperatures electrons have enough energy to redistribute more evenly among the valleys and the effect disappears.},\cite{Canali_PRB75}. Therefore, at elevated temperatures the time-of-flight of conduction electrons across the Si channel of a transport device drops monotonically with increasing electric field. If electrons are initially spin polarized, then the accepted Elliott-Yafet spin relaxation theory suggests that the spin depolarization during transport is dependent only on the time-of-flight. In this theory the spin and momentum relaxation times are proportional \cite{Elliott_PR54,Yafet_1963}, so the resulting spin polarization increases with electron drift velocity. Indeed, we have confirmed this expectation in previous experiments where low and moderate applied fields ($<$1~kV/cm) were used in studying the extraordinarily long spin lifetime \cite{Huang_PRL07,Huang_PRB10}, and in demonstrating spin injection and detection in ferromagnet-silicon hybrid systems \cite{Appelbaum_Nature07,Huang_APL07}.

In this Letter, we experimentally demonstrate an unexpected dependence of the spin polarization on the electric field in silicon at $T\geq$30~K in high electric fields. With increasing field, the spin polarization of detected electrons first increases as expected from the Elliott-Yafet static lifetime model; however, above 2~kV/cm it starts to decrease, showing a Gunn-effect dependence akin to a negative differential spin lifetime without any simultaneous negative differential charge/spin mobility. The origin of this counterintuitive behavior is then elucidated by Monte Carlo simulations and a quantitative analytical description. When the electron ensemble is out of thermal equilibrium with the lattice, an efficient spin relaxation mechanism becomes accessible due to field-induced intervalley scattering. We quantify the spin relaxation time as a function of both the lattice and electron ensemble temperature. The latter provides a means to determine the dependence of spin relaxation in silicon on the electric field and will enable the optimization of spintronics devices.

Coherent spin precession and spin valve measurements were performed to observe the nonequilibrium depolarization effect and to quantify the negative differential region of spin lifetime. In both experiments, we employed all-electrical devices in which spin-polarized electrons (aligned with the in-plane magnetization direction of a ferromagnetic thin-film source) are tunnel injected through a Schottky metal contact and into a 225 $\mu$m -thick wafer of nominally undoped Si(100). The electrons then drift across the wafer thickness due to an applied electric field, and are collected by a second ferromagnetic film where their spin is analyzed using a ballistic spin detection scheme. The results presented below do not depend on the injection and detection techniques but only on the spin and charge transport characteristics of the Si channel. We therefore include all device-specific description in the supplemental material \cite{SUPPLMAT}, and refer the interested reader to Refs. \cite{Huang_PRL07,Appelbaum_Nature07,Huang_APL07} for further details.

\begin{figure}
\includegraphics[width=8.6cm]{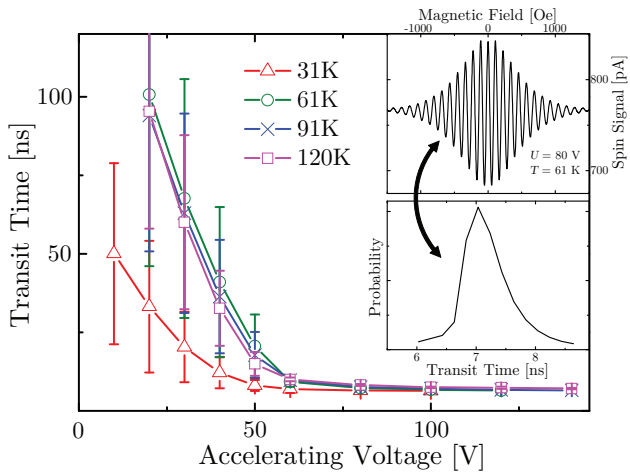}
\caption{\label{FIG1}
Average transit time across a 225-micron-thick intrinsic silicon device as a function of the applied voltage for various temperatures. Error bars indicate the transit time uncertainty (extracted from the width of the time-of-flight distribution; see lower inset). Top inset: Symmetrized spin precession data at $T=61$~K and 80~V (3.5~kV/cm). The data show high spin coherence with well-defined oscillation field period. Lower inset: spin current transit time distribution obtained by transforming the precession signal (see text).}
\end{figure}

The time-of-flight distribution of the electron current can be recovered from quasistatic spin precession measurements by applying an external magnetic field, $B$, perpendicular to the injected spin direction but parallel to the electric field \footnote{The applied magnetic field is oriented perpendicular to in-plane magnetizations of the ferromagnetic source and detector films. These magnetization directions are not affected by the field due to their shape anisotropy.}. This magnetic field induces spin precession at frequency $\omega=g\mu_B B/\hbar$, where $g$ is the electron g-factor, $\mu_B$ is the Bohr magneton, and $\hbar$ is the reduced Planck constant. We denote the time-of-flight distribution by $D(t)$ where its mean and standard deviation are, respectively, measures of the average transit time and of diffusion and dephasing effects in the channel. The signal contribution from the spin component parallel to the detector magnetization of an electron arriving at the detector in the time interval [$t,t+dt$] is therefore $D(t)\cos{\omega t}dt$.  The variation in quasistatic detected signal is then $D(\omega)\propto \int_0^\infty D(t) \cos{\omega t}dt$; by repeating the measurement at various applied magnetic fields one can map the precession frequency dependence of the detected signal \cite{Huang_PRB10}. Finally, the empirical time-of-flight distribution is recovered without any model fitting by the inverse Fourier transform of $D(\omega)$. An example of this transformation between $D(\omega)$ and $D(t)$ is shown in the coupled insets to Fig.~\ref{FIG1}. The main figure shows the average transit time across the silicon channel as a function of applied voltage for several temperatures. Clearly, increasing the internal electric field with applied voltage reduces the transit time until the onset of velocity saturation for voltages $\apprge 60$~V (electric field $\sim2.7$~kV/cm) \cite{Canali_PRB75}. It is important to note that no negative differential mobility is seen.

\begin{figure}
\includegraphics[width=8.6cm]{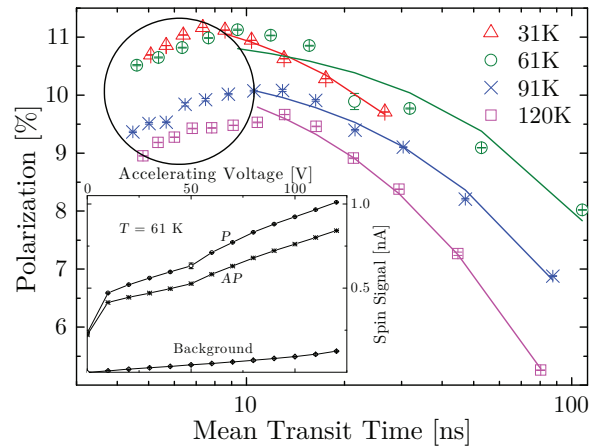}
\caption{\label{FIG2} Measured current polarization in a ferromagnet/silicon/ferromagnet device as a function of transit time in the silicon channel. Solid lines indicate exponential fits to the low-field (long transit time) data and indicate long spin lifetimes in that regime. Effects of spin depolarization from electric-field-induced spin relaxation are evident at high-fields (short transit time; circled data). Inset: Example spectroscopy at 61~K showing spin signal in a parallel (P) and antiparallel (AP) magnetic configuration, with background signal for subtraction.
}
\end{figure}

To measure the spin polarization of collected electrons, we have performed independent spin-valve measurements as a function of the electric field. Here, a small external magnetic field of $\approx$~20 Oe is applied along the source magnetization axis and thus no spin precession is induced. The final spin polarization after transport is extracted by the ratio $\mathcal{P} = (I_P-I_{AP})/(I_P+I_{AP})$, where $I_P$ is the measured signal current in a configuration where the in-plane injector and detector magnetization directions (and hence spin initialization and measurement axes) are parallel, and $I_{AP}$ is for antiparallel configuration. The inset in Fig.~2(a) shows an example of this spectroscopy taken by interleaving $P$ and $AP$ measurements at each applied voltage to avoid signal drift from field-induced stress. We also include the background detector current taken under conditions of zero injection current after the signal measurement. It has subsequently been subtracted in the polarization calculation to avoid misinterpreting a spurious dilution for spin depolarization.

Figure \ref{FIG2}(a) shows the measured polarization as a function of the average transit time ($\tau_{tr}$).  This ratio depends on the spin relaxation time in the Si channel by $(I_P-I_{AP})/(I_P+I_{AP}) =\mathcal{P}_0e^{-\tau_{tr}/\tau_s}$ where  $\mathcal{P}_0$ (limited by the spin-injection and detection efficiencies of the device) is the optimal attainable value. The figure shows that at long transit times, the polarization increases with reducing the transit time, as expected from the Elliott-Yafet theory. However, at short transit times (circled data) the trend is unexpectedly opposite. This observation of a nonmonotonic spin polarization Gunn effect is the main experimental result of this Letter.

The origin of this phenomenon is a transition to a previously-ignored regime where electric field directly enables a spin relaxation pathway. The field-induced momentum relaxation enhancement, as implied by the saturation in charge transport data of Fig.~1, is not commensurate with the spin relaxation enhancement. Applying the accepted Elliott-Yafet theory (proportionality of spin and momentum relaxation times) would therefore lead to the false conclusion that the rising polarization with initially increasing transit time is indicative of an unphysical \emph{negative} spin lifetime.

\begin{figure}
   \centering
   \includegraphics[width=8.6cm]{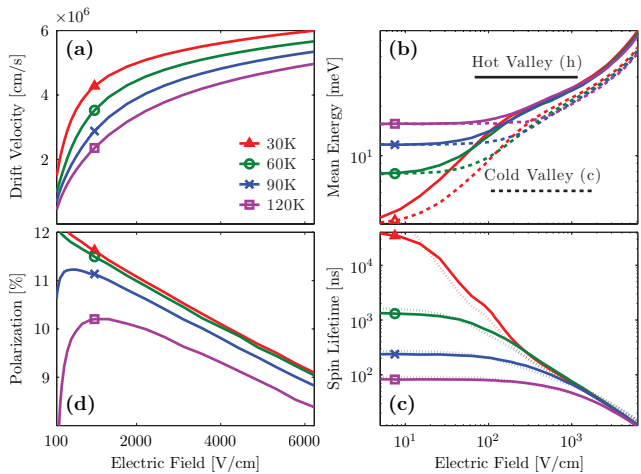}
   \caption{(a) Electron drift velocity, (b) mean energy, (c) spin relaxation time and (d) final polarization as a function of electric field, calculated from numerical integration of the distribution obtained from Monte Carlo simulation. The dotted lines in (c) denote Eq.~(1).}
   \label{MC}
\end{figure}

We have performed Monte-Carlo simulations in order to elucidate the charge transport and spin relaxation of conduction electrons heated by the electric field (``hot'' electrons). A full description of the numerical procedure is provided in the supplemental material \cite{SUPPLMAT}, and here we summarize the important features. Ellipsoidal energy bands are used to model the equivalent six conduction band valleys \cite{Herring_PR57}. Momentum relaxation mechanisms are modeled by electron-phonon interactions (both intravalley and intervalley processes) and intravalley electron scattering from ionized impurities \footnote{An impurity density of 10$^{12}$~cm$^{-3}$ is used to match the resistivity of the silicon wafers (4~k$\Omega$$\cdot$cm at room-temperature). At this low density, the transport parameters are practically independent on impurity scattering at high electric fields.}. Between scattering events, electrons are treated as classical particles accelerated by the electric field. Typically, an out-of-equilibrium electron distribution reaches its steady-state within $1$~ns regardless of the initial condition. Figure~\ref{MC}(a)-(b) show the corresponding drift velocity and mean energy as a function of applied electric field. Hot (cold) valleys refer to the four (two) valleys whose axis is perpendicular to (collinear with) the electric field. The mean energy in a hot (cold) valley is higher (lower) due to the different projections of electric field on the ellipsoidal energy bands.

The spin relaxation due to electron-phonon interactions is calculated by integration of intravalley and intervalley spin-flip matrix elements \cite{Li_PRL11} while using the Monte-Carlo hot-electron distributions. The solid lines in Fig.~\ref{MC}(c) show the results of this numerical procedure. Note that spin relaxation due to scattering with ionized impurities is negligible in nearly intrinsic wafers. As a result the total spin lifetime in our devices is significantly longer than in heavily doped Si channels \cite{Sasaki_APL10,Sasaki_APL11,Ando_APL11}.
Figure~\ref{MC}(d) shows the spin polarization $\mathcal{P}_0\exp(-\tau_{tr}/\tau_s)$, where $\mathcal{P}_0=0.125$ is chosen to fit the experimental injection and detection efficiencies. The average transit time $\tau_{tr}$ is calculated from the drift velocity after transport across 225~$\mu$m. At low fields, the polarization rises with electric field since the increase of drift velocity surpasses the decrease of spin relaxation time. As the drift velocity begins to saturate in high fields, the polarization drops slowly due to the enhanced reduction of the spin relaxation time. This dependence of spin polarization on the electric field agrees well with the experimental results and reproduces the Gunn-type behavior (here shown as a function of the field).

\begin{figure}
   \centering
   \includegraphics[width=8.6cm]{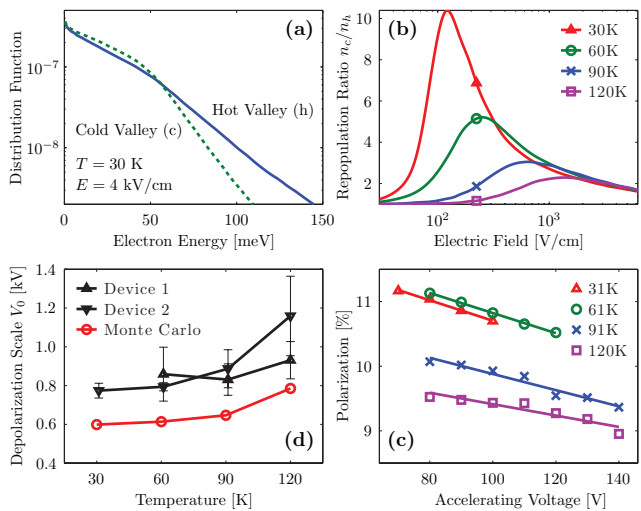}
   \caption{(a) Electron distributions in hot (solid line) and cold (dashed line) valleys. The total electron density is 10$^{12}$~cm$^{-2}$, the electric field is 4~kV/cm and the lattice temperature is 30~K. (b) Ratio between electron densities in cold and hot valleys as a function of the field. (c) Experimental depolarization at high-fields (extracted from Fig.~(2)). (d)  Characteristic scale of the drop in spin polarization as a function of temperature.}
   \label{final}
\end{figure}

We focus on the important underlying spin relaxation mechanism and analytically quantify the observed effect. From the mean energy (Fig.~\ref{MC}(b)), one can see that electrons driven by the electric field become hot enough to undergo intervalley electron-phonon processes during which the electron delivers to the lattice a few tens of meV \cite{Canali_PRB75}. We consider $f$-processes at which electrons are scattered between valleys of different crystal axes. This process dominates the spin relaxation of hot electrons since it involves a direct coupling of valence and conduction bands \cite{Li_PRL11}. To conserve crystal momentum, the phonon wavevector resides on the $\Sigma$ axis. The symmetry-allowed phonon modes for spin relaxation are $\Sigma_1$ and $\Sigma_3$ with respective phonon energies of $\Omega_{f,1}\approx 47$~meV and $\Omega_{f,3}\approx 23$~meV. The $\Sigma_3$ mode allows for scattering between all valleys and the $\Sigma_1$ mode restricts them to the case that one of the involved valley axes is not perpendicular to the spin quantization axis. To analytically quantify the spin relaxation we functionalize the hot electron distributions. Fig.~\ref{final}(a) shows the Monte-Carlo steady-state energy distributions in hot and cold valleys at 30~K and 4~kV/cm. The distribution (in each of the valleys) can be described by a two-component heated Boltzmann distribution. At the low energy part, the effective temperature of the electron distribution can be extracted from the mean energy $\frac{3}{2}k_BT_e$, shown in Fig.~\ref{MC}(b). At the higher energy part, intervalley process tend to cool the system. To simplify the analysis below we employ an effective electron temperature, $T'_e=T+\gamma(T_e-T)$  where $T$ is the lattice temperature and $\gamma\approx0.9$ is a constant that mimics the cooling effect due to intervalley scattering at high electron energies \cite{SUPPLMAT}. We denote effective lattice and electron parameters by $y_i=\Omega_{f,i}/k_BT$, $y'_{i,\mu}=\Omega_{f,i}/k_BT'_{e,\mu}$ where $i$ denotes the phonon modes and $T'_{e,\mu}$ is the effective temperature of the electrons in a cold ($\mu=c$) or a hot ($\mu=h$) valley. Using the above, we arrive at an analytical spin lifetime \cite{SUPPLMAT},
\begin{equation}
\frac{1}{\tau_{s}}\approx\mathcal{C}\!\!\sum^{\mu=h,c}_{i=1,3}\!\!A_{i,\mu}n_{\mu}\frac{\exp(y_i\!-\!y'_{i,\mu})\!+\!1}{\exp(y_i)\!-\!1}(\frac{4}{3}y'^{-\frac{1}{2}}_{i,\mu}\!+\!\sqrt{2}),\label{eq:taus}
\end{equation}
where $\mathcal{C}=0.036~$ns$^{-1}$ is a constant related to the spin-orbit coupling parameter of the $X$ point at the edge of the Brillouin zone. $A_{1,h}\!=\!8\,(12)$, $A_{3,h}\!=\!1.5\,(1.25)$, $A_{1,c}\!=\!8\,(4)$ and $A_{3,c}\!=\!0.5\,(0.75)$ are symmetry related parameters when the electric field is collinear with (perpendicular to) the spin-quantization axis. The $n_c$ and $n_h$ denote, respectively, the fractional population at cold and hot valleys where $2n_c+4n_h=1$. Figure~\ref{final}(b) shows the repopulation ratio, $n_c/n_h$. The asymmetry in valley population is largest at intermediate fields since electrons that reside in hot valleys become energetic enough for intervalley scattering to cold valleys. At high fields, scattering in the opposite direction also becomes accessible and the valley population is more symmetrical ($n_c/n_h \approx$~2 in this regime). Substituting the extracted values of $n_{\mu}$ and $T'_{e,\mu}$ in Eq.~(\ref{eq:taus}) reproduces the spin relaxation of hot electrons as can be seen from the comparison between the dotted and solid lines in Fig.~\ref{MC}(c). At equilibrium conditions where $T'_{e,\mu}=T$ this mechanism is greatly suppressed (especially at low temperatures).

We can quantitatively compare the results of our calculations with empirical data by extracting a characteristic voltage scale $V_0$ for $f$-process-induced spin depolarization, where we approximate $\mathcal{P}\approx \mathcal{P}_0(1-\frac{V}{V_0})$ in the high electric-field regime. Fits to the measured spin polarization data are shown in Fig. \ref{final}(c), and the consistent temperature dependence of $V_0$ for several devices is shown in Fig. \ref{final}(d). The Monte-Carlo prediction extracted from the high-field regime in Fig.~\ref{MC}(d) closely resembles the empirical values in both magnitude and lattice-temperature dependence due to the more efficient generation of intervalley scattering (and hence lower $V_0$) at lower temperatures. By taking the high-field limit of Eq.~(\ref{eq:taus}), one can write $ V_0 \approx v_d/[d (\tau_s^{-1})/dE] \approx 0.7$~kV where $v_d$ is the saturated drift velocity. This close correspondence confirms our interpretation of field-induced $f$-process spin depolarization in the experiment, in a regime where acoustic-phonon-mediated scattering as well as scattering with states at the spin hot-spot \cite{Li_PRL11,FABIANWU} are too small to account for this effect \cite{SUPPLMAT}.

Finally, we note that a recent theoretical proposal suggesting that stochastic polarization fluctuations can be amplified by spin-dependent mobility \cite{Qi_PRL06} has also been termed a ``spin Gunn effect''. Our experiment and theory differ from this scheme in that the mobility and diffusion constants are spin independent, electron-electron collisions are negligible \cite{Damico_EPL01}, and the physical origin of the effect is attributed to the signature of spin-orbit coupling on electron-phonon intervalley scattering. The phenomenon observed here, as in its charge-based counterpart, is due to a strong electric field-induced relaxation which leads to a qualitatively different spin transport regime distinct from expectations based on the Elliott-Yafet theory.

In conclusion, high electric fields present in silicon devices can substantially change the dominant physical mechanism of spin relaxation. In this regime, the Elliott-Yafet mechanism mediated by intravalley acoustic phonons is far outweighed by the depolarizing effects of inelastic scattering with intervalley $f$-process phonons created by the efforts of the system to recover thermal equilibrium. This behavior is expected to be critically important in the design of devices making use of spins to transmit information, especially when strong static electric fields are required \cite{Dery_APL11}. Similarly, the derived dependence of the spin lifetime on the electric field is of fundamental importance to the design of semiconductor devices that make use of spin as an alternative degree of freedom \cite{Dery_Nature07,LOU,Fabian_APS07,Behin_NatureNano10}.

Work at UMD is supported by the Office of Naval Research and the National Science Foundation. We acknowledge the support of the Maryland NanoCenter and its FabLab. Work at UR is supported by AFOSR and NSF (No. FA9550-09-1-0493 and No. DMR 1124601).

\end{document}